\documentclass[]{spie}

\usepackage{amsmath,amsfonts,amssymb}
\usepackage{graphicx}
\usepackage[colorlinks=true,allcolors=blue]{hyperref}

\usepackage{dcolumn}
\usepackage{bm}
\usepackage{mathtools}
\usepackage{esint}
\usepackage{xcolor}
\usepackage{dsfont}
\usepackage[outline]{contour}
\usepackage{nicefrac}
\usepackage{multirow}

\newcommand{\bea}{\begin{eqnarray}}
\newcommand{\eea}{\end{eqnarray}}

\newcommand{\phd}{\phantom{\dag}}
\newcommand{\ph}{\phantom{.}}

\newcommand{\noi}{\noindent}
\newcommand{\no}{\nonumber}

\title{New mechanisms to engineer magnetic skyrmions and topological superconductors}

\author[a]{Panagiotis Kotetes}
\author[b]{Yun-Peng Huang}
\affil[a]{CAS Key Laboratory of Theoretical Physics, Institute of Theoretical Physics, Chinese Academy of Sciences, Beijing 100190, China}
\affil[b]{Beijing National Laboratory for Condensed Matter Physics, and Institute of Physics, Chinese Academy of Sciences, Beijing 100190, China}
\authorinfo{Further author information: (Send correspondence to Panagiotis Kotetes and Yun-Peng Huang)\\
Panagiotis Kotetes: E-mail: kotetes@itp.ac.cn\\
Yun-Peng Huang: E-mail: huangyunpeng@iphy.ac.cn}

\begin{document} 
\maketitle

\begin{abstract}
We propose an alternative route to stabilize magnetic skyrmions which does not require Dzyaloshinkii-Moriya interactions, magnetic anisotropy, or an external Zeeman field. Our so-called magnetic skyrmion catalysis (MSC) solely relies on the emergence of flux in the system's ground state. We review scenarios that allow for a nonzero flux and summarize the magnetic skyrmion phases that it induces. Among these, we focus on the so-called skyrmionic spin-whirl crystal (Sk-SWC$_4$) phase. We discuss aspects of MSC using a concrete model for topological superconductivity, which describes the surface states of a topological crystalline insulator in the presence of proximity induced pairing. By assuming that the surface states can exhibit the Sk-SWC$_4$ phase, we detail how the addition of a pairing gap generates a chiral superconductor. For this purpose, we construct a low-energy model which renders the mechanism for topological superconductivity transparent. Moreover, by employing this model, we perform a self-consistent investigation of the appearance of the Sk-SWC$_4$ phase for different values of the pairing gap and the ground state's flux. Our analysis verifies the catalytic nature of our mechanism in stabilizing the Sk-SWC$_4$ phase, since the magnetization modulus becomes enhanced upon ramping up the flux. The involvement of MSC further shields magnetism against the suppression induced by the pairing gap. Remarkably, even if the pairing gap fully suppresses the Sk-SWC$_4$ phase for a given value of flux, this skyrmion phase can be restored by further increasing the flux. Our findings demonstrate that MSC enables topological superconductivity in a minimal and robust fashion.
\end{abstract}

\keywords{Magnetic Textures, Magnetic Skyrmion Crystal, Magnetic Skyrmion Catalysis, Topological Superconductor, Majorana Fermion, Topological Quantum Computing}

\vskip 1cm

\section{Introduction}\label{Sec:Intro}

Being in a position to controllably generate and tune the coupling between the electron's spin and its charge gives access to a variety of intriguing magnetoelectric phenomena~\cite{Edelstein1990,VolovikBook,SHE,Linder,Banerjee}. The conventional mechanism for spin-orbit coupling (SOC) originates from the Zeeman effect. This coupling has a relativistic origin and dictates all electronic systems, since it does not require the violation of spatial inversion (${\cal I}$) symmetry. Notably, breaking ${\cal I}$-symmetry opens the door to further SOC mechanisms~\cite{Winkler}, e.g., the Dresselhaus and Rashba effects. The former arises in bulk systems which lack a center of inversion. In contrast, the Rashba SOC is of an extrinsic nature and appears due to the so-called structural ${\cal I}$ asymmetry. This emerges when different type of materials interface each other, thus creating a nonuniform electrostatic enviroment and a nonzero electric field throughout the structure. In particular, when the system feels the presence of a uniform electric field $\bm{E}$, an additional term $\big(\bm{E}\times\bm{s}\big)\cdot\bm{p}$ contributes to the energy of the electron and couples its spin $\bm{s}$ to its momentum $\bm{p}$. The appearance of this term reflects the presence of an additional velocity $\propto\bm{E}\times\bm{s}$, which allows the electron to couple to electric fields which are oriented perpedicular to its motion, even in the absence of an external magnetic field. Such an unconventional behavior constitutes one of the pillars for spintronic applications~\cite{SpintronicsRev} and dissipationless spin transport~\cite{SHERMP}.

The Rashba SOC is typically engineered in a two-dimensional electron gas (2DEG) which becomes confined in a ``slice" of a multilayer three-dimensional (3D) quantum well structure. The arising electric field, which breaks inversion symmetry and in turn traps the 2DEG, is tailored by carefully choosing the materials for the various layers. In general, Rashba SOC is ubiquitous at interfaces and terminations of systems possessing strong atomic SOC. More recently, a Rashba-type SOC has been shown to govern the helical surface states of 3D topological insulators (TIs)~\cite{HasanKane,QiZhang}, thus, providing a natural playground for magnetoelectric phenomena when an additional uniform magnetic field is applied to the system. Among these, one also finds the possibility of realizing parity anomaly~\cite{HasanKane,QiZhang,Ion,Nomura,Tse,Drew,Mogi,Hano}. Moreover, the emergence of such an ${\cal I}$-symmetry breaking SOC on each TI surface opens perspectives for engineering topological superconductivity~\cite{FuKane}. Specifically, when a pairing gap is introduced to the system, Majorana fermion quasiparticles can be trapped at superconducting vortices, thus laying the foundations for building a topological quantum computer~\cite{Kitaev2003,Nayak2008}.

The broad range of phenomena in which the Rashba SOC finds application highlights the importance of finding suitable systems that can naturally harbour it, and reflects the strong need to discover ways to artificially synthesize it. One promising route to generate Rashba-type SOC, even in centrosymmetric materials, is using magnetic textures. These correspond to spatially nonuniform configurations of the magnetization field $\bm{M}(\bm{r})$ defined in coordinate space $\bm{r}$. Magnetic textures are unique objects, in the sense that they lead to the violation of time-reversal (${\cal T}$) symmetry (TRS) and at the same time they are able to generate SOC which is odd under ${\cal I}$, akin to the Rashba SOC~\cite{MostovoyFerro,BrauneckerSOC,Choy}. Hence, they can effectively produce situations of coexisting Rashba SOC and uniform magnetic fields and open the door to the various phenomena mentioned in the previous paragraphs and others~\cite{Yokoyama,Hals2,Christensen_18}. Finally, the simultaneous presence of these two symmetry breaking effects renders them ideal ingredients for engineering Majorana quasiparticles~\cite{VolovikBook,ReadGreen,Volovik99,KitaevChain,HasanKane,QiZhang,FlensbergOppenStern,Zlotnikov,KotetesTSCchapter,KovalevRev}.

Magnetic textures can induce topological superconductivity in two distinct ways. In the first possibility, one relies on magnetic textures crystals (MTCs), which constitute magnetic textures that repeat periodically in space. In many instances, a MTC can be described in term of a small number of Fourier components $\bm{M}_{\bm{Q}}$, hence giving rise to a so-called multi-$\bm{Q}$ profile~\cite{MartinBatista,HayamiRev,HuangZhouKotetes,Takagi,hayami2023chern}. Up to date, there exists a plethora of theoretical works which have demonstrated the manners in which interfacing MTCs with spin-singlet superconductivity can give rise to topological superconductivity in both 1D~\cite{KarstenNoSOC,Ivar,KlinovajaGraphene,NadgPerge,KotetesClassi,Nakosai,Selftuned1,Selftuned2,Selftuned3,Pientka1,Ojanen,Pientka2} and 2D~\cite{KotetesClassi,Nakosai,Mendler,WeiChen,MorrSkyrmions,MorrTripleQ,SteffensenPRR}, respectively. On the experimental side, MTCs with a nontrivial spatial profile in 1D have been realized in hybrid devices consisting of carbon nanotubes and nanomagnets~\cite{Kontos}. More recently, scanning tunne\-ling microscopy (STM) measurements have demonstrated the formation of helical MTCs in Fe chains deposited on top of Re(0001). Remarkably, the same experiments have also provided strong evidences for Majorana zero modes (MZMs) when Re becomes a superconductor~\cite{KimWiesendanger}. Further, the coexistence of conventional superconductivity and magnetic skyrmion type of MTCs, which have a nontrivial winding in 2D, has also been experimentally demonstrated in 2D Fe/Ir magnets on Re(0001)~\cite{Kubetzka}. The second pathway to harness the coupling between superconductivity and magnetic textures with the goal of creating Majorana quasiparticles, is when the textures can be viewed as extended zero-dimensional (0D) defects. The most prominent candidate for such a magnetic texture defect is a magnetic skyrmion~\cite{BogdanovSkMagMet,Muelhbauer,Tokura1,Tokura2,Heinze,Kawamura,LeonovMostovoy,HayamiZeroField,Wulfhekel,TripleQWiesendanger}, which can be controllably introduced in chiral ferromagnets~\cite{Bogdanov,BogdanovHubert,NagaosaTokuraSkX,BogdanovPanagopoulos}. Theoretical works have demonstrated that a pair of MZMs can be trapped at the core and rim of a magnetic skyrmion defect~\cite{BalatskySkyrmion,Yang2016,MikeSkyrmion,Kovalev,Rex2019,Garnier2019a,Garnier2019b} when the latter is coupled to a conventional superconductor. Important experimental advancements along these lines were recently made in [IrFeCoPt]/Nb hybrids~\cite{PanagopoulosSkX}, where composite excitations consisting of magnetic skyrmions and (anti)vortices were experimentally observed~\cite{PanagopoulosSkyrmionVortex}.

In the above cutting edge experimental efforts, the engineering of magnetic skyrmions is pursued using a well established route~\cite{Bogdanov,BogdanovHubert}, which relies on chiral magnets in the presence of Dzyaloshinkii-Moriya interaction (DMI)~\cite{Dzyaloshinskii,Moriya}, out-of-plane magnetic anisotropy, and an out-of-plane magnetic field. In spite of the fact that magnetic skyrmion textures, both crystals and defects, have been experimentally created using this approach in a number of instances, following the same path for engineering topological superconductivity requires more caution. First of all, some of the materials in which magnetic skyrmions have been observed are rather inhomogeneous and disordered, which is harmful for the stability of Majorana excitations and their manipulation~\cite{Kitaev2003,Nayak2008}. Second, the currently employed mechanism requires the application of a magnetic field which does not generally bode well with spin-singlet superconductivity. As a result, this sets strong limitations on the possible combination of magnets and superconductors which can be used in tandem for the successful engineering of magnetic skyrmions. Finally, when it comes to realizing a quantum computer it is most desired to control the qubits using electrostatic means since this promises an enhanced degree of tunability and in principle also opens the door to scalable solid state structures. Therefore, it would be beneficial if the magnetic host could demonstrate insulating behavior and at the same time be responsive to electrostatic gating.

In this Manuscript, we discuss aspects of an alternative route to creating magnetic skyrmions and topological superconductivity, which appears in principle less demanding than the so far pursued approach. The mechanism in question has been recently proposed by the present authors in Ref.~\citenum{Huang} and was coined magnetic skyrmion catalysis (MSC). In essence, within the MSC framework, one wishes to identify a source field for skyrmion charge, which will generate magnetic skyrmions in an analogous manner to which a magnetic field induces magnetization. Inducing magnetic skyrmion textures by means of MSC is free from all the requirements that the currently established approach for magnetic skyrmion imposes, i.e., DMI interactions, magnetic anisotropy, and a magnetic field. The only requirement is that the electronic degrees of freedom which lead to magnetism feel a nonzero flux. Relaxing the various constraints for generating magnetic textures promises to unfold an alternative pool of candidates for engineering magnetic skyrmions and topological superconductivity. Prominent materials for observing MSC are magnets which in the paramagnetic phase behave as doped Chern insulators, or, doped semimetals which contain topologically protected band touchings. When these systems are additionally of semiconducting nature, one expects electrostatic tunability to be in principle accessible in the low density regime, in analogy to the situation taking place in currently accessible semiconductor-superconductor hybrid devices~\cite{Vuik,Antipov,Woods,Mikkelsen}. Moreover, by choosing antiferromagnetic type of systems with a low or zero net magnetization, promises to lead to an improved proximity effect between the magnet and the superconducting element.

A number of fundamental aspects of the mechanism of MSC have already been examined in our previous work. However, a number of open questions still remain. Here, we present a self-consistent analysis of the phenomenon, and identify the windows of stability of the magnetic skyrmion order as a function of the flux of the ground state and the strength of the pairing gap of the system. In Ref.~\citenum{Huang} we discussed two possibilities within which skyrmionic MTCs can emerge by violating TRS in tetragonal magnets. These two scenarios consider as a starting point materials which exhibit the so-called spin-vortex crystal (SVC) and fourfold-symmetric spin-whirl skyrmion crystal (SWC$_4$) phases as magnetic ground states~\cite{Christensen_18}. These two magnetic phases have been theoretically predicted to be realizable in doped Fe-based superconductors~\cite{Christensen_18}. For the demonstration of MSC in Ref.~\citenum{Huang}, we considered as a model system the surface of a crystalline TI, which harbors spinful topologically protected modes which feature an energy spectrum that contains a single quadratic band crossing~\cite{FuTCI}. For doped surface states which additionally exhibit tendency to develop spin-density wave order, the so-called skyrmion variants of the SVC and SWC$_4$, i.e., the Sk-SVC and Sk-SWC$_4$ phases, become accessible when TRS is violated. The latter takes place when an energy gap opens and lifts the degeneracy of the topological band touching point. When the Sk-SVC (Sk-SWC$_4$) magnetic order coexists with conventional superconductivity, a chiral topological superconductor becomes effectively created which harbors two (one or two) chiral Majorana modes when an edge is introduced to the system. In the present work, we further elaborate on the scenario of Sk-SWC$_4$ order coexisting with superconductivity. We derive a low energy model for the arising topological superconductor. By employing this model we obtain the self-consistency equation dictating the emergence of the Sk-SWC$_4$. We find that for a sufficiently strong interaction in the magnetic channel, the system is self-tuned to a topological superconductor ground state. Moreover, this phase remains robust in a quite wide parameter regime which gives hope that it could be observable in a realistic material.

Our manuscript is organized as follows. In Sec.~\ref{Sec:SectionII} we discuss a number of properties of skyrmion MTCs and explain the rationale behind proposing the MSC mechanism. In addition, we provide a conceptual comparison between MSC and the usual DMI approach. In Sec.~\ref{Sec:SectionIII} we discuss the properties of the concrete model that we study in this work. This model describes a spinful doped Chern insulator, which leads to a Chern number that is equal to two units. In Sec.~\ref{Sec:SectionIV} we remind the reader of the phenomenological Landau-level description of MSC for tetragonal magnets and introduce the Sk-SVC and Sk-SWC$_4$ MTCs. Next, in Sec.~\ref{Sec:SectionV}, we consider the emergence of topological superconductivity by assuming the coexistence of conventional superconductivity with the Sk-SWC$_4$ phase. We detail how the presence of the MTC gaps out the various flat segments of the Fermi surface, which is assumed to exhibit strong nesting features, thus promoting the stabilization of spin-density waves. For this demonstration, we derive a effective low-energy model which allows us to describe the key effects of the Sk-SWC$_4$ ground state. Section~\ref{Sec:SectionVI} contains the main analysis of this work which regards the self-consistent magnetic phase diagram, as well as the emergence and stability of the magnetic skyrmion ground state. The self-consistent approach is particularly important, especially in the presence of a pairing gap, since magnetism and spin-singlet superconductivity are generally competing in nature. Finally, Sec.~\ref{Sec:SectionVII} discusses our conclusions and provides an outlook.

\section{Mechanisms for Magnetic Texture Crystals:\\ DMI Approach vs Magnetic Skyrmion Catalysis}\label{Sec:SectionII}

In this section we wish to explain the rationale behind proposing the mechanism of MSC. Moreover, we wish to conceptually compare it with the standard DMI-based route to obtain magnetic skyrmions. In order to do this, it is first important to discuss a few general properties of magnetic textures.

As mentioned earlier, apart from violating TRS, magnetic textures have attracted significant interest because they can also simultaneously induce Rashba-type SOC. For this, magnetic textures need to violate ${\cal I}$ symmetry. For a magnetic defect this is automatically satisfied, while for a multi-$\bm{Q}$ type of MTC, the magnetic vectors at which magnetism sets in should be incommensurate to the lattice. Assuming that these conditions are satisfied, the effective SOC generated by a magnetic texture contributes at lowest order in momenta with an energy term of the general form $W_{ij}s_ip_j$ with the $W_{ij}$ coefficients given by the expression:
\begin{align}
W_{ij}\propto\int d^3x\,\varepsilon_{ik\ell}M_k(\bm{r})\frac{\partial M_\ell(\bm{r})}{\partial x_j}\end{align}

\noi where we introduced the fully-antisymmetric Levi-Civita symbol and assumed the repeated index summation convention. The various indices run over the labels of the spatial coordinates of the position vector $\bm{r}=(x_1,x_2,x_3)$. For example, a helical MTC of the form:
\begin{align}
\bm{M}(\bm{r})=\big(M_{||}\sin(Qx_1),0,M_\perp\cos(Qx_1)\big)
\end{align}

\noi leads to $W_{21}\propto M_{||}M_{\perp}$ with the remaining $W_{ij}$ components to be zero. As a result, the induced SOC is of the form $p_1s_2$. Notably, since the magnetic helic crystal is periodic, we can further introduce the winding vectors for the magnetization field~\cite{Christensen_18}:
\begin{align}
\bm{w}_j=\frac{1}{2\pi}\int_{\rm MUC}dx_j\,\hat{\bm{M}}(\bm{r})\times \frac{\partial\hat{\bm{M}}(\bm{r})}{\partial x_j}\end{align}

\noi where the integration is over the magnetic unit cell (MUC) and $\hat{\bm{M}}(\bm{r})=\bm{M}(\bm{r})/|\bm{M}(\bm{r})|$. Note that no repeated index summation is assumed above. As long as the magnetization field of the MTC depends only on a single coordinate, then $\bm{w}_i$ has components which take integer ($\mathbb{Z}$) values and its orientation yields the direction of the electron's spin which becomes coupled to the $i$-th component of the momentum. The above imply that a helical MTC is a topological object, i.e., smooth deformations which satisfy $|\bm{M}(\bm{r})|\neq0$ cannot unwind it. Note that the above winding vector can be also defined for nonperiodic magnetization profiles, as long as the integration is over a closed path, i.e., a path where at its beginning and end points the magnetization is the same.

As a natural extension, we now consider the following noncollinear MTC profile:
\begin{align}
\bm{M}(\bm{r})=\big(M_{||}\sin(Qx_1),0,M_\perp\cos(Qx_1)\big)+(0,M_{||}\sin(Qx_2),M_\perp\cos(Qx_2)\big)
\end{align}

\noi which corresponds to the SWC$_4$ profile~\cite{Christensen_18}. Besides the nonzero element $W_{21}$ that we found previously for one helix, here, we observe that the MTC renders one more element nonzero, i.e., $W_{12}\propto-M_{||}M_{\perp}$. From the above, we find that the SWC$_4$ crystal effectively induces the term $p_1s_2-p_2s_1$ in the electron's Hamiltonian, which is no other than the celebrated Rashba SOC term arising when an electric field builds up out-of-the $x_1$-$x_2$ plane. Interestingly, while the structure of the SWC$_4$ resembles to a spin skyrmion crystal, a skyrmion charge cannot be defined for the SWC$_4$. This is because $|\bm{M}(\bm{r})|=0$ for certain position vectors $\bm{r}$ in the MUC. Nevertheless, removing these zeros renders the MTC skyrmionic. For instance, adding a uniform magnetization component $(0,0,B)$ leads to $|\bm{M}(\bm{r})|\neq0$ for all coordinate vectors $\bm{r}$. As a result, the MTC has a nonzero skyrmion charge for $|B|<2|M_\perp|$.

A skyrmionic MTC defined in the $x_1$-$x_2$ plane is dictated by a nonzero topological charge which we here denote $\Phi_3$. This charge is defined by the expression:
\begin{align}
\Phi_3=\frac{1}{4\pi}\int_{\rm MUC}dx_1dx_2\,\hat{\bm{M}}(\bm{r})\cdot\left[\frac{\partial\hat{\bm{M}}(\bm{r})}{\partial x_1}\times\frac{\partial\hat{\bm{M}}(\bm{r})}{\partial x_2}\right],
\end{align}

\noi where similar to the case of the winding vectors, the area integral in the present case has to be taken over a closed/compact domain in the event that the magnetic texture is not periodic. The topological charge $\Phi_3$ is quantized and take $\mathbb{Z}$ values~\cite{VolovikBook}. It corresponds to the flux threaded in the MUC due to the magnetic skyrmion crystal. Note that for a MTC defined in three dimensions, one can further define the components $\Phi_1$ and $\Phi_2$ which correspond to the fluxes threaded through the remaining planes.

The nonzero winding vectors and the skyrmion charge are defining for a skyrmion MTC, and any system which effectively induces either one, or ideally both of these, favors the stabilization of skyrmionic magnetic textures. Indeed in the proposal based on DMI, the Rashba SOC dictating the electronic system induces a term in the energy functional describing the magnetization, which is of the form $e_{ij}W_{ij}$. Hence, adding a SOC which breaks ${\cal I}$ symmetry generates the tensor elements $e_{ij}$ which act as sources for the winding elements $W_{ij}$ of the magnetization. However, since the Rashba SOC respects TRS there is no term source for $\Phi_3$. This is exactly the reason for the external magnetic field. This violates TRS and in the additional presence of the nonzero $e_{ij}$, it indirectly sources $\Phi_3$. The fact that DMI and an external field are both required in order to device a source field for skyrmion charge is in fact the biggest difference compared to the MSC that we propose.

Within the MSC framework, the presence of a nonzero flux $\vartheta$ in the ground state plays the role of a source field for $\Phi_3$, and contributes with an additional term of the form$-\vartheta{\cal C}$ to the magnetic energy of the system where we defined:
\begin{align}
{\cal C}=\frac{1}{4\pi}\int dx_1dx_2\ph \bm{M}(\bm{r})\cdot\left[\frac{\partial\bm{M}(\bm{r})}{\partial x_1}\times\frac{\partial\bm{M}(\bm{r})}{\partial x_2}\right]\,.\label{eq:SkDensity}
\end{align}

\noi The quantity ${\cal C}$ and the skyrmion charge $\Phi_3$ become nonzero simultaneously. Therefore, the flux $\vartheta$ also acts as a source for $\Phi_3$. Three-spin interactions similar to ${\cal C}$ have been previously employed at a Hamiltonian level~\cite{Paramekanti2017,JiangJiang,HuangSheng,Paramekanti} to phenomenologically describe the emergence of chiral ground states in magnets. Here, we associate the flux $\vartheta$, which acts as a source field for ${\cal C}$, with the nontrivial topology of the magnet in its nonordered phase. In particular, for a doped Chern insulator $\vartheta$ can be linked to the Chern number of the bands, while for a doped topological semimetal with the order of the band touching point. See also the work in Ref.~\citenum{Levitov} for a related proposal, which explores how to generate smooth skyrmion textures on top of a ferromagnetic ground state in doped Chern systems.

Magnetic skyrmion excitations have long ago been theoretically predicted for~\cite{Sondhi,Fertig} and have subsequently been experimentally captured~\cite{Barrett} in the close cousin of the Chern insulator, i.e., the quantum Hall system. In the latter, the nonzero Chern number appears for the bandstructure only in the presence of an external so-called quantizing magnetic field which leads to Landau levels.

\section{Chern Metals and Induced Flux}\label{Sec:SectionIII}

The essence of the mechanism of MSC is to consider a magnetic system which is characterized by a nonzero flux $\vartheta$ in its ground state. A paradigmatic model which satisfies this requirement is the following concrete Hamiltonian:
\begin{align}
\hat{h}(\bm{p})=\frac{p_1p_2}{m_1}\kappa_1+m\kappa_2+\frac{p_1^2-p_2^2}{2m_2}\kappa_3=\left(\frac{p_1p_2}{m_1},m,\frac{p_1^2-p_2^2}{2m_2}\right)\cdot\bm{\kappa}\equiv\bm{d}(\bm{p})\cdot\bm{\kappa}\,,\label{eq:Chern}
\end{align}

\noi where we employed the Pauli matrices $\kappa_{1,2,3}$ which act in a space spanned by an abstract ``orbital" degree of freedom, which is not required to be specified for our present analysis. We remark that the above Hamiltonian describes either spin species of the electron, since we assume that the system in the nonmagnetic phase is invariant under the full group of SU(2) rotations in spin space.

When $m=0$, the energy spectrum of the remaining Hamiltonian $\hat{h}_0(\bm{p})\equiv\hat{h}_{m=0}(\bm{p})$ yields a single quadratic band touching point at $\bm{p}=\bm{0}$. The emergence of a single quadratic band crossing can take place either when dealing with a particular region of the bandstructure of a crystalline material, or, with a single surface of a topological crystalline insulator~\cite{FuTCI}. The fact that the above Hamiltonian alone fails to describe a bulk crystalline system is reflected in the respective Chern number obtained for $m\neq0$, which is equal to a single unit and possesses a sign determined by the sign of the gap $m$. Given that the band touching point is of second order and thus carries vorticity of two units, we find that the respective Chern number is equal to half of the vorticity and thus is fractionally quantized. Such a behavior is anomalous in the sense that it cannot characterize a bulk 2D crystalline TI which is invariant under fourfold rotations.

In Ref.~\citenum{Huang} we focused on Chern systems, such as the one in Eq.~\eqref{eq:Chern}, and we obtained the contribution $F^{(3)}$ to the magnetic energy which is cubic with respect to the components of the magnetization $\bm{M}(\bm{r})$. This term is obtained under the condition that the magnetization plays the role of a perturbation. Such a perturbative approach is meaningful when studying the magnetic instability of the system. In view of the potential realization of skyrmionic MTCs with multi-$\bm{Q}$ configurations, it is convenient to decomposed the magnetization field in terms of Fourier components which are given by the relation $\bm{M}(\bm{q})=\int d\bm{r}\, e^{-i\bm{q}\cdot\bm{r}}\bm{M}(\bm{r})$. TRS violation due to a nonzero $\vartheta$ allows for a cubic contribution in the expansion which contains the term ${\cal L}(\bm{q},\bm{p})=\bm{M}(-\bm{q}-\bm{p})\cdot\big[\bm{M}(\bm{q})\times\bm{M}(\bm{p})\big]$, which relates to the skyrmion charge density. Specifically, the emergence of ${\cal L}(\bm{q},\bm{p})$ leads to the energy term:
\begin{align}
F^{(3)}=\frac{1}{4\pi}\int\frac{d\bm{q}}{(2\pi)^2}\int\frac{d\bm{p}}{(2\pi)^2}\ph\vartheta(\bm{q},\bm{p})\ph{\cal L}(\bm{q},\bm{p})\,,\label{eq:Fcubic}
\end{align}

\noi where we introduced the flux density $\vartheta(\bm{q},\bm{p})$. In the limit $\bm{q}\rightarrow\bm{0}$ and $\bm{p}\rightarrow\bm{0}$ Eq.~\eqref{eq:Fcubic} yields Eq.~\eqref{eq:SkDensity} and $\vartheta(\bm{q},\bm{p})$ becomes ``replaced" by $\vartheta$. The analysis of Ref.~\citenum{Huang} obtained that in the ferromagnetic instability limit, where $F^{(3)}\mapsto-\vartheta{\cal C}$, the flux coefficient reads as:
\begin{align}
\vartheta=\sum_{s=0,1,2}\frac{4}{(s+1)!s!}
\sum_{\alpha}\int\frac{d\bm{p}}{\pi}\ph
i\varepsilon_{znm}\left<\partial_{p_n}\bm{u}_\alpha(\bm{p})\right|\big[\mathds{1}-\hat{{\cal P}}_{\alpha}(\bm{p})\big]
\big[\varepsilon_{\alpha}(\bm{p})-\hat{h}(\bm{p})\big]^{s-2}
\left|\partial_{p_m}\bm{u}_\alpha(\bm{p})\right>\partial_\mu^s f[\varepsilon_{\alpha}(\bm{p})-\mu],\label{eq:theta}
\end{align}

\noi where $\left|\bm{u}_{\alpha}(\bm{p})\right>$ are the eigenvectors of $\hat{h}(\bm{p})$ with corresponding energy disper\-sions $\varepsilon_\alpha(\bm{p})$ and projectors $\hat{{\cal P}}_{\alpha}(\bm{p})=\left|\bm{u}_{\alpha}(\bm{p})\right>\left<\bm{u}_{\alpha}(\bm{p})\right|$. The function $f(\epsilon-\mu)$ appearing above corresponds to the Fermi-Dirac distribution defined in the grand canonical ensemble and eva\-lua\-ted at an energy $\epsilon$ for a chemical potential $\mu$. Due to the presence of the projectors, we infer that $\vartheta$ stems from interband-only transitions~\cite{Huang}, which is typical of quantities that have a topological origin. The expression for $\vartheta$ simplifies for two-band models such as the one of interest, and takes the compact form:
\begin{align}
\vartheta=\sum_{s}^{0,1,2}\sum_\alpha^{{\color{black}\pm1}}\int\frac{d\bm{p}}{\pi}\ph\frac{(-2)^s\alpha^{1+s}\Omega(\bm{p})}{(s+1)!s![\varepsilon(\bm{p})]^{2-s}}\ph\partial_\mu^sf[-\alpha\varepsilon(\bm{p})-\mu]\label{eq:Theta2Band}
\end{align}

\noi where the quantity:
\begin{align}
\Omega(\bm{p})=\frac{1}{2}\ph\hat{\bm{d}}(\bm{p})\cdot\Big[\partial_{p_1}\hat{\bm{d}}(\bm{p})\times\partial_{p_2}\hat{\bm{d}}(\bm{p})\Big]\,
\end{align}

\noi denotes the Berry curvature of the valence band~\cite{Niu}, with $\hat{\bm{d}}(\bm{p})=\bm{d}(\bm{p})/|\bm{d}(\bm{p})|$, $\varepsilon_\pm(\bm{p})=\pm\varepsilon(\bm{p})$ and $\varepsilon(\bm{p})=|\bm{d}(\bm{p})|$.

\section{Magnetic ground states in the presence of flux}\label{Sec:SectionIV}

After reviewing the result regarding the induced flux in the ground state due to a topologically nontrivial band structure, we proceed with analyzing the qualitative consequences of the cubic term on the magnetic phase diagram when the magnetic ordering can be described by a small number of Fourier components. We carry out this investigation using a Landau-type of formulation. For concreteness, we restrict our discussion to 2D doped Chern insulators with tetragonal symmetry. Our approach can be extended to other point group symmetries along the same lines. Note that our discussion in this section closely follows Ref.~\citenum{Huang}.

\begin{figure}[t!]
\centering
\includegraphics[width=\textwidth]{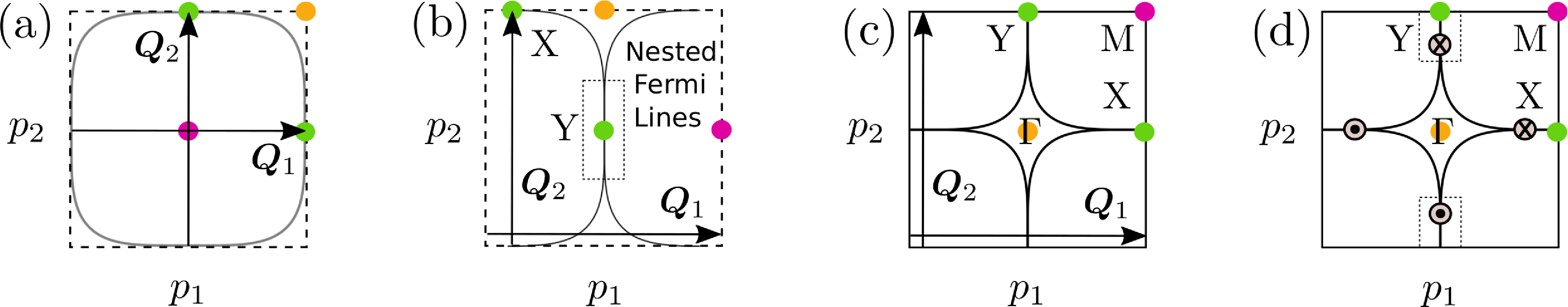}\vspace{0.1in}
\caption{\label{fig:Figure1}
(a) Fermi surface for the model of Eq.~\eqref{eq:Chern} with $m_1=1$, $m_2=0.7$, $m=0.038$ and $\mu=0.082$. One observes the formation of flat Fermi lines which are nested with $\bm{Q}_{1,2}$ which have a modulus $Q=|\bm{Q}_{1,2}|=2\sqrt{2m_2\sqrt{\mu^2-m^2}}$. (b) Fermi surface after downfolding to the magnetic Brillouin zone (MBZ) for the nesting vector $\bm{Q}_1$. The points X and Y of the MBZ originate from points of the Fermi surface experiencing single-$\bm{Q}$ nesting with $\bm{Q}_{1,2}$, respectively. This MBZ constitutes corresponds to the Brilouin zone used for the low-energy model of Sec.~\ref{Sec:SectionV}. The low-energy model focuses on a given pair of nested flat Fermi lines and in particular on the one shown to be enclosed by a rectangular box. (c) Fermi surface after downfolding to the magnetic Brillouin zone (MBZ) for the nesting vectors $\bm{Q}_1$ and $\bm{Q}_2$. Note that the center of coordinates point of the original Brillouin zone in (a) becomes the M point in the MBZ. The ${\rm \Gamma}$ point of the MBZ (orange dot) stems from a point in (a) which lies energetically away from zero, and experiences nesting with $\bm{Q}_\pm$. (d) Here, we depict two pairs of point nodes with opposide vorticity which arise for the case discussed in (a) but with zero flux, i.e., $m=0$, when additionally, the SWC$_4$ phase coexists with a conventional superconducting gap. Since the flux is zero, TRS is preserved and the total charge of the nodes is zero. The nodes are gapped out upon switching on even an infinitesimally weak flux which induces the $\bm{M}_{\bm{Q}_\pm}$ components of the magnetization field.}
\end{figure}

We focus on a system which exhibits tendency to develop magnetic ground states at the star of the ordering wave vectors $\pm\bm{Q}_{1,2}$, with $\bm{Q}_1=Q(1,0)$ and $\bm{Q}_2=Q(0,1)$. For an illustration, see Fig.~\ref{fig:Figure1} where we employ the model of Eq.~\eqref{eq:Chern} to demonstrate a situation where the Fermi surface of the system exhibits strong nesting features. In the presence of the nonzero flux, the cubic term also introduces ordering at the additional wave vectors $\bm{Q}_{\pm}=\bm{Q}_1\pm\bm{Q}_2$. However, magnetic ordering for the star $\bm{Q}_{\pm}$ generally takes place at a different critical temperature than for star $\pm\bm{Q}_{1,2}$. Here, we assume that the magnetic instability at $\bm{Q}_{1,2}$ happens at a higher temperature and thus ordering at this star drives magnetism. Hence, the effects of the additional star can be treated in a perturbative fashion.

The order parameters which correspond to the primary star $\bm{Q}_{1,2}$ are described by the Landau-type energy functional:~\cite{Christensen_18}
\bea
F_{\bm{M}_{1,2}}&=&
\alpha\big(|\bm{M}_1|^2+|\bm{M}_2|^2\big)+\frac{\tilde{\beta}}{2}\big(|\bm{M}_1|^2+|\bm{M}_2|^2\big)^2
+\frac{\beta-\tilde{\beta}}{2}\big(|\bm{M}_1^2|^2+|\bm{M}_2^2|^2\big)+(g-\tilde{\beta})|\bm{M}_1|^2|\bm{M}_2|^2\nonumber\\
&+&\frac{\tilde{g}}{2}\big(|\bm{M}_1\cdot\bm{M}_2|^2+|\bm{M}_1\cdot\bm{M}_2^{\ast}|^2\big)\,.\label{eq:F12}
\eea

\noi All the accessible magnetic ground states of the above have been precisely identified in Ref.~\citenum{Christensen_18}. Among the possible phases one also finds the SWC$_4$, which is characterized by the order parameters:
\begin{align}
\bm{M}_{1}=M\big(i\cos\lambda,0,\sin\lambda\big)\qquad{\rm and}\qquad \bm{M}_{2}=M\big(0,i\cos\lambda,\sin\lambda\big)\,,\label{eq:M12}
\end{align}

\noi where $\lambda$ and $M$ are known parameters which depend on the coefficients of the various terms appearing in the Landau functional.

In the presence of the cubic term, the energy functional is required to be extended in order to include the emergent $\bm{M}_{\bm{Q}_\pm}\equiv\bm{M}_\pm$ components. Since these components are not thermodynamically critical, we can restrict to the lowest order contributions which can describe their appearance. Thus, we arrive to the energy functional:
\bea
F_{\bm{M}_{1,2,\pm}}=F_{\bm{M}_{1,2}}+\bar{\alpha}\left(|\bm{M}_+|^2+|\bm{M}_-|^2\right)-\vartheta_{1,2}\big[
\bm{M}_{+}^*\cdot\big(\bm{M}_{1}\times\bm{M}_{2}\big)-\bm{M}_{-}^*\cdot\big(\bm{M}_1\times\bm{M}_2^*\big)+{\rm c.c.}\big],
\label{eq:free_energy}
\eea

\noi where $\vartheta_{1,2}$ corresponds to $\vartheta(\bm{q},\bm{p})$ after being eva\-lua\-ted for all permutations of the momenta $\bm{q},\bm{p}=\big\{\bm{Q}_1,\bm{Q}_2,\bm{Q}_\pm\big\}$. The Euler-Lagrange equations for $\bm{M}_\pm$ directly provide their expression in terms of $\bm{M}_{1,2}$ and after Ref.~\citenum{Huang}, we write:
\begin{align}
\bm{M}_+=+\frac{\vartheta_{1,2}}{\bar{\alpha}}\bm{M}_1\times\bm{M}_2\phd\,{\rm and}\,\phd\bm{M}_-=-\frac{\vartheta_{1,2}}{\bar{\alpha}}\bm{M}_1\times\bm{M}_2^*.\label{eq:Mpm}
\end{align}

\noi Plugging the above back into Eq.~\eqref{eq:free_energy}, we find an effective free energy depending only on $\bm{M}_{1,2}$:
\begin{align}
{\cal F}_{\bm{M}_{1,2}}=F_{\bm{M}_{1,2}}-\frac{\gamma}{2}\left(|\bm{M}_1\times\bm{M}_2|^2+|\bm{M}_1\times\bm{M}_2^*|^2\right)\,,\label{eq:NewF12}
\end{align}

\noi where we introduced the variable
\begin{align}
\gamma=\frac{2\vartheta_{1,2}^2}{\bar{\alpha}}\geq0\,.
\end{align}

One observes that the emergence of MSC in square lattice systems relies on a fluctuations type of mechanism, which is mediated by the noncritical order parameters $\bm{M}_\pm$. This is a peculiarity of the given crystal symmetry group, since it does not support simultaneous magnetic ordering at three or more wave vectors which add up to the null vector. In addition, it is worth noting that although ${\cal F}_{\bm{M}_{1,2}}$ alone respects TRS, magnetic skyrmion crystal ground states are now possible thanks to the additional presence of $\bm{M}_\pm$. The modified functional for $\bm{M}_{1,2}$, which encodes the feedback of the $\bm{M}_\pm$ components can be also precisely mi\-ni\-mi\-zed and it was shown in Ref.~\citenum{Huang} that it shares the same type of minima for $\bm{M}_{1,2}$ with Eq.~\eqref{eq:F12}. However, the energies for these ground states differ due to the appearance of the additional $\bm{M}_\pm$ star. The modification of the energies can be equivalently introduced in Eq.~\eqref{eq:F12} by means of the shifted coefficients:
\begin{align}
g\mapsto g-\gamma\quad{\rm and}\quad\tilde{g}\mapsto\tilde{g}+\gamma\,.
\end{align}

\noi In the case of the SWC$_4$ of interest, we find that the additional components take the following form:
\begin{align}
\bm{M}_{\pm}=\tilde{M}\cos\lambda\big(i\sin\lambda,\pm i\sin\lambda,\cos\lambda\big)\,,
\end{align}

\noi where $\tilde{M}=\theta M$ with $\theta=-\vartheta_{1,2}/\bar{\alpha}$. We find that star\-ting from a SWC$_4$ ground state for $\gamma=0$ with $\bm{M}_\pm=\bm{0}$, can induce the skyrmionic Sk-SWC$_4$ phase with a skyrmion charge $|\Phi_3|=1,2$ when $\gamma$ is switched on. This is because $\bm{M}_\pm$ are now nonzero and modify $\bm{M}(\bm{r})$. See also Ref.~\citenum{Huang} for further details.

\section{Low-Energy Model for Topological Superconductivity}\label{Sec:SectionV}

We now proceed with the investigation of the properties of a Sk-SWC$_4$ ground state in the additional presence of a conventional pairing gap which can be intrinsic or induced by means of the proximity to a bulk superconductor. Our starting point is the model of Eq.~\eqref{eq:Chern}. Assuming that magnetism is of an itinerant origin and driven by Fermi surface nesting, the properties of the topological superconductor are mainly determined by the band which gives rise to the Fermi surface. Note that this holds as long as the pairing gap $\Delta\geq0$ is much smaller than the Fermi energy denoted $E_F$. Therefore, to study the Fermi-surface magnetic instability and the emergence of topological superconductivity one can simplify the model by restricting to the following energy dispersion:
\begin{align}
\varepsilon(\bm{p})=\sqrt{\big(p_1p_2/m_1\big)^2+\big[\big(p_1^2-p_2^2\big)/2m_2\big]^2+m^2}-\mu\,,
\end{align}

\noi which gives rise to the Fermi surface of the system assuming that $\mu>0$. Notably, an additional approximation can be made in the regime $|m|\ll|\mu|$, i.e., when the gap at the band touching point is sufficiently away from the Fermi level. In this case, one can drop the term $m\kappa_2$ of the nonsuperconducting Hamiltonian and instead take into account the effects of $m$ by means of the following: (i) observe that the presence of $m$ effectively renormalizes $\mu$, so that it becomes $\mu\rightarrow\sqrt{\mu^2-m^2}$, and (ii) introduce the order parameters $\bm{M}_\pm$ which implicitly depend on $\vartheta$ and, in turn, on the mass gap $m$. Hence, in the remainder, we restrict to the study of the following approximate energy dispersion:
\begin{align}
\varepsilon(\bm{p})\approx\sqrt{\big(p_1p_2/m_1\big)^2+\big[\big(p_1^2-p_2^2\big)/2m_2\big]^2}-\sqrt{\mu^2-m^2}\,.\label{eq:ChernApprox}
\end{align}

We now extend our formalism to second quantization so to include superconductivity more conveniently. By employing the multicomponent spinor
\begin{align}
\bm{\Psi}^\dag(\bm{r})=\big(\psi_\uparrow^\dag(\bm{r}),\,\psi_\downarrow^\dag(\bm{r}),\,\psi_\downarrow(\bm{r}),\,-\psi_\uparrow(\bm{r})\big),
\end{align}

\noi we find that the Hamiltonian operator describing the system in second quantization reads as:
\begin{align}
\hat{H}=\frac{1}{2}\int d\bm{r}\ph\bm{\Psi}^\dag(\bm{r})\hat{H}_{\rm BdG}(\hat{\bm{p}},\bm{r})\bm{\Psi}(\bm{r}).
\end{align}

\noi The above is expressed in terms of the so-called Bogoliubov - de Gennes (BdG) ``single-particle" Hamiltonian:
\begin{align}
\hat{H}_{\rm BdG}(\hat{\bm{p}},\bm{r})=\varepsilon(\hat{\bm{p}})\tau_3+\Delta\tau_1+\bm{M}(\bm{r})\cdot\bm{\sigma}\,,
\end{align}

\noi where $\varepsilon(\hat{\bm{p}})$ is obtained from Eq.~\eqref{eq:ChernApprox} after upgrading the momentum to the respective operator $\hat{\bm{p}}=-i\bm{\nabla}$. The BdG Hamiltonian is expressed using Kronecker products defined in Nambu and spin space. To represent the matrices, we employ the spin (Nambu) Pauli matrices $\sigma_{1,2,3}$ ($\tau_{1,2,3}$) along with the respective unit matrix $\mathds{1}_\sigma$ ($\mathds{1}_\tau$). To simplify the notation, unit matrices are neglected throughout unless it is deemed necessary for avoiding confusion.

Since we are mainly interested in studying the nesting-driven magnetic instability, we primary focus on the regions enclosing the flat parts of the Fermi surface which are expected to be the ones primarily contributing to the magnetic condensation energy. Therefore, we can focus on momenta with $|\bm{p}|\sim p_F$. By virtue of the fourfold rotational symmetry which is preserved by the Sk-SWC$_4$ magnetic order, it is sufficient to investigate the properties of only a pair of nested surfaces. The Hamiltonian for the other can be obtained in a similar fashion.

To proceed, we here choose to study the Fermi surface near $p_1=\pm p_F$. For this reason it is best to focus on right/left movers with momenta primarily concentrated near $p_1=\pm p_F$. With an eye to an expansion about these $x_1$ axis Fermi momenta, it is convenient to decompose the field operator according to the expression:
\begin{align}
\bm{\Psi}(\bm{r})=e^{+i\bm{Q}_1\cdot\bm{r}/2}\bm{\Psi}_R(\bm{r})+e^{-i\bm{Q}_1\cdot\bm{r}/2}\bm{\Psi}_L(\bm{r})\,.
\end{align}

\noi In the vicinity of these momenta, the dispersion shows a very weak dispersion with $p_2$ due to the ``flatness" of the Fermi sheets which is crucial for nesting. See Fig.~\ref{fig:Figure1}(b) for an illustration. By introducing the mover basis through the enlarged spinor $\widetilde{\bm{\Psi}}^\dag(\bm{r})=\big(\bm{\Psi}_R^\dag(\bm{r}),\,\bm{\Psi}_L^\dag(\bm{r})\big)
$, we end up with the fol\-lowing expressions:
\begin{align}
\hat{\cal H}_{\rm BdG}(\hat{\bm{p}},\bm{r})=
\varepsilon_-(\hat{\bm{p}})\rho_3\tau_3+\mathds{1}_\rho\big[\varepsilon_+(\hat{\bm{p}})\tau_3+\Delta\tau_1\big]+\sum_{\bm{Q}_s=\pm\bm{Q}_{1,2,\pm}}\left(\mathds{1}_\rho+\sum_{\nu=\pm}e^{-\nu i\bm{Q}_1\cdot\bm{r}}\rho_\nu\right)e^{i\bm{Q}_s\cdot\bm{r}}\bm{M}_{\bm{Q}_s}\cdot\bm{\sigma}
\label{eq:MoversHam}
\end{align}

\noi where the Pauli matrices $\rho_{1,2,3}$, their associated unit matrix $\mathds{1}_\rho$ and raising/lowering operators $\rho_\pm=(\rho_1\pm i\rho_2)/2$, all act in mover space. By further proceeding with restricting to momenta near $p_1=\pm p_F$, we find that $\varepsilon_-(\bm{p})\approx\pm\upsilon_F\hat{p}_1$ for right/left movers and $\varepsilon_+(\bm{p})\approx- p_2^2/2m_\perp$, with $m_\perp$ an effective mass for tranverse motion. Since the Fermi surface is flat, $m_\perp$ is expected to be large. Using the model of Eq.~\eqref{eq:ChernApprox}, we obtain that:
\begin{align}
\upsilon_F=\frac{p_F}{m_2}\qquad {\rm and} \qquad\frac{m_2}{m_\perp}=1-2\left(\frac{m_2}{m_1}\right)^2.
\end{align}

We now concentrate on the magnetic part. The sum with respect to the wave vectors runs over the following set $\pm\{\bm{Q}_{1,2},\bm{Q}_\pm\}$. Within the spirit of this rotating-wave-approximation (RWA) the spatial dependence carried by $\bm{M}_{\bm{Q}_1}$ becomes fully eliminated. This however does not happen for the contributions obtained from the last term in Eq.~\eqref{eq:MoversHam}. Indeed, the respective terms retain their nontrivial spatial dependence but nevertheless also become modified in this new frame. Most interestingly, they get ``mixed", i.e., a Fourier component with wave vector $\bm{Q}_s$ generates additional phase factors to the usual $e^{i\bm{Q}_s\cdot\bm{r}}$. Nonetheless, within the spirit of the RWA, terms which are also modulated in the spatial direction denoted by $\bm{Q}_1$ need to be also dropped. In the present case, this implies that terms $\propto e^{i\bm{Q}_\pm\cdot\bm{r}}$ need to be discarded. Hence, based on the above arguments, we obtain:
\begin{align}
\hat{\cal H}_{\rm BdG}(\hat{\bm{p}},\bm{r})=
\upsilon_F\hat{p}_1\rho_3\tau_3+\mathds{1}_\rho\left(\Delta\tau_1-\frac{\hat{p}_2^2}{2m_\perp}\tau_3\right)
+\Big[\bm{M}_1\rho_++e^{i\bm{Q}_2\cdot\bm{r}}\big(\bm{M}_2\mathds{1}_\rho+\bm{M}_+\rho_++\bm{M}_-^*\rho_-\big)+{\rm h.c.}\Big]\cdot\bm{\sigma}.\label{eq:MoversHamNice}
\end{align}

We proceed by treating the modulated term in the $x_2$ direction ($||\bm{Q}_2$) within second order perturbation theory. Such an approach is most conveniently demonstrated by transferring to momentum space and truncating the resul\-ting Hamiltonian down to the lowest order of momentum transfers. We first define the uniform part of the above Hamiltonian in momentum space as:
\begin{align}
\hat{\cal H}_0(\bm{p})=\upsilon_Fp_1\rho_3\tau_3+\mathds{1}_\rho\left(\Delta\tau_1-\frac{p_2^2}{2m_\perp}\tau_3\right)+\bm{M}_1\cdot\rho_+\bm{\sigma}+\bm{M}_1^*\cdot\rho_-\bm{\sigma}\,.
\end{align}

\noi Second, we express the perturbation in a compact form accor\-ding to $e^{i\bm{Q}_2\cdot\bm{r}}\hat{\cal V}+{\rm h.c.}$ where we defined the Hamiltonian term $\hat{\cal V}=\big(\bm{M}_2\mathds{1}_\rho+\bm{M}_+\rho_++\bm{M}_-^*\rho_-\big)\cdot\bm{\sigma}$. With the help of the above steps we are now in a position to write the truncated equations for lowest order coupled modes which read as:
\bea
\hat{\cal V}\widetilde{\bm{\Phi}}(\bm{p})+\hat{\cal H}_0(\bm{p}+\bm{Q}_2)\widetilde{\bm{\Phi}}(\bm{p}+\bm{Q}_2)&=&E\widetilde{\bm{\Phi}}(\bm{p}+\bm{Q}_2)\,,\\
\hat{\cal V}\widetilde{\bm{\Phi}}(\bm{p}-\bm{Q}_2)+\hat{\cal V}^\dag\widetilde{\bm{\Phi}}(\bm{p}+\bm{Q}_2)+\hat{\cal H}_0(\bm{p})\widetilde{\bm{\Phi}}(\bm{p})&=&E\widetilde{\bm{\Phi}}(\bm{p})\,,\\
\hat{\cal V}^\dag\widetilde{\bm{\Phi}}(\bm{p})+\hat{\cal H}_0(\bm{p}-\bm{Q}_2)\widetilde{\bm{\Phi}}(\bm{p}-\bm{Q}_2)&=&E\widetilde{\bm{\Phi}}(\bm{p}-\bm{Q}_2)\,.
\eea

\noi We remark that the above have been obtained strictly by truncating additional terms which also in principle couple to the modes with vectors $\widetilde{\bm{\Phi}}(\bm{p}\pm\bm{Q}_2)$. Since here we aim at inferring the effects of the perturbation on the $\widetilde{\bm{\Phi}}(\bm{p})$ vector, which is considered to be an approximate eigenvector of $\hat{\cal H}_0(\bm{p})$, we have:
\begin{align}
\widetilde{\bm{\Phi}}(\bm{p}+\bm{Q}_2)\approx \hat{G}_+(\bm{p})\hat{\cal V}\widetilde{\bm{\Phi}}(\bm{p})\qquad{\rm and}\qquad
\widetilde{\bm{\Phi}}(\bm{p}-\bm{Q}_2)\approx \hat{G}_-(\bm{p})\hat{\cal V}^\dag\widetilde{\bm{\Phi}}(\bm{p})
\end{align}

\noi where we introduced the operators:
\begin{align}
\hat{G}_\pm(\bm{p})=\frac{1}{\hat{\cal H}_0(\bm{p})-\hat{\cal H}_0(\bm{p}\pm\bm{Q}_2)}=\frac{\tau_3}{E_{F_\perp}\pm \upsilon_{F_\perp}p_2}\,,
\end{align}

\noi with $E_{F_\perp}=(m_2/m_\perp)E_F\ll E_F$, $\upsilon_{F_\perp}=p_F/m_\perp=(m_2/m_\perp)\upsilon_F\ll\upsilon_F$, and $E_F=p_F^2/(2m_2)$. In the present case, we are mainly interested in small $p_2$, in which limit the above functions approximately become:
 \begin{align}
\hat{G}_\pm(\bm{p})\approx\left(1\mp \frac{2p_2}{p_F}\right)\frac{\tau_3}{E_{F_\perp}}\,.
\end{align}

From the above, we immediately obtain that the Hamiltonian $\hat{\cal H}_0(\bm{p})$ becomes perturbed and results in the expression $\hat{\cal H}_{\rm eff}(\bm{p})\approx\hat{\cal H}_0(\bm{p})+\delta\hat{\cal H}_0(\bm{p})$ where $\delta\hat{\cal H}_0(\bm{p})=\hat{\cal V}^\dag\hat{G}_+(\bm{p})\hat{\cal V}+\hat{\cal V}\hat{G}_-(\bm{p})\hat{\cal V}^\dag$. Given the above conditions we restrict to $|p_2|\ll p_F$ and obtain that:
\begin{align}
\delta\hat{\cal H}_0(\bm{p})=\left(\big\{\hat{\cal V},\hat{\cal V}^\dag\big\}+\big[\hat{\cal V},\hat{\cal V}^\dag\big]\frac{2p_2}{p_F}\right)\frac{\tau_3}{E_{F_\perp}}\,.
\end{align}

\noi The above can be further simplified by compactly re-expressing the perturbation in the form $\hat{\cal V}\equiv\hat{\bm{V}}\cdot\bm{\sigma}$. Under this condition, we obtain:
\bea
\delta\hat{\cal H}_0(\bm{p})&=&
\left(\big\{\hat{V}_n,\hat{V}_n^\dag\big\}\mathds{1}_\sigma+i\varepsilon_{nm\ell}\big[\hat{V}_n,\hat{V}_m^\dag\big]\sigma_\ell\right)\frac{\tau_3}{E_{F_\perp}}+
\left(\big[\hat{V}_n,\hat{V}_n^\dag\big]\mathds{1}_\sigma+i\varepsilon_{nm\ell}\big\{\hat{V}_n,\hat{V}_m^\dag\big\}\sigma_\ell\right)\frac{2p_2}{p_F}\frac{\tau_3}{E_{F_\perp}}
\,,\no\\
&=&
+\frac{2M^2+\tilde{M}^2\cos^2\lambda\big[3-\cos(2\lambda)\big]}{E_{F_\perp}}\tau_3+\frac{4M\tilde{M}\cos\lambda\sin(2\lambda)}{E_{F_\perp}}\rho_1\tau_3+\frac{2\tilde{M}^2\cos^2\lambda\sin(2\lambda)}{E_{F_\perp}}\rho_3\tau_3\sigma_2\no\\
&&-\frac{4\big(M^2+\tilde{M}^2\cos^2\lambda\big)\sin(2\lambda)}{E_{F_\perp}}\frac{p_2}{p_F}\tau_3\sigma_1-\frac{4M\tilde{M}\cos\lambda}{E_{F_\perp}}\frac{p_2}{p_F}\rho_1\tau_3\sigma_1-\frac{4M\tilde{M}\cos\lambda\sin(2\lambda)}{E_{F_\perp}}\frac{p_2}{p_F}\rho_2\tau_3\sigma_3\no\\
\eea

\noi where in the first row we employed the Einstein convention for repeated index summation.

In order to better understand the effect of the above contribution we return to $\hat{\cal H}_0(\bm{p})$. Since $\bm{M}_1=M(i\cos\lambda,0,\sin\lambda)$, we obtain the concrete expression:
\begin{align}
\hat{\cal H}_0(\bm{p})=\upsilon_Fp_1\rho_3\tau_3+\mathds{1}_\rho\left(\Delta\tau_1-\frac{p_2^2}{2m_\perp}\tau_3\right)+M\sin\lambda\rho_1\sigma_3-M\cos\lambda\rho_2\sigma_1\,.
\end{align}

\noi At this stage, we perform two back-to-back unitary transformations. The first is effected by $(\rho_2+\rho_3)/\sqrt{2}$ and the second by $(\rho_2\sigma_2+\sigma_3)/\sqrt{2}$. These lead to:
\begin{align}
\hat{\cal H}_0'(\bm{p})=\upsilon_Fp_1\rho_2\tau_3+\mathds{1}_\rho\left(\Delta\tau_1-\frac{p_2^2}{2m_\perp}\tau_3\right)-M\rho_1\big(\cos\lambda+\sin\lambda\sigma_3\big)\,.
\end{align}

\noi As we observe, the above transformation block dia\-go\-na\-li\-zes the bare Hamiltonian and splits it into two $\sigma_3=\pm1$ blocks. By means of carrying out the same transformation on the correction to the Hamiltonian $\delta\hat{\cal H}_0(\bm{p})$, we find:
\bea
\delta\hat{\cal H}_0'(\bm{p})&=&\frac{2M^2+\tilde{M}^2\cos^2\lambda\big[3-\cos(2\lambda)\big]}{E_{F_\perp}}\tau_3-\frac{4M\tilde{M}\cos\lambda\sin(2\lambda)}{E_{F_\perp}}\rho_3\tau_3\sigma_1+\frac{2\tilde{M}^2\cos^2\lambda\sin(2\lambda)}{E_{F_\perp}}\tau_3\sigma_3\no\\
&+&\frac{4\big(M^2+\tilde{M}^2\cos^2\lambda\big)\sin(2\lambda)}{E_{F_\perp}}\frac{p_2}{p_F}\tau_3\sigma_1-\frac{4M\tilde{M}\cos\lambda}{E_{F_\perp}}\frac{p_2}{p_F}\rho_3\tau_3-\frac{4M\tilde{M}\cos\lambda\sin(2\lambda)}{E_{F_\perp}}\frac{p_2}{p_F}\rho_3\tau_3\sigma_3
\,.\qquad
\eea

\noi We proceed by projecting onto the $\sigma_3=\pm1$ sectors. After an additional unitary transformation effected by $(\rho_1-\rho_3)/\sqrt{2}$, we end up with the effective Hamiltonians:
\begin{align}
\hat{\cal H}_{\rm eff}^\sigma(\bm{p})=-\left(\frac{p_2^2}{2m_\perp}-\mu_\sigma+\upsilon_Fp_1\rho_2-\upsilon_{F_{\perp,\sigma}}p_2\rho_1\right)\tau_3+\Delta\tau_1+M_\sigma\rho_3\,,
\end{align}

\noi where we introduced the effective block dependent magnetizations, chemical potential and transverse Fermi velocities, respectively:
\bea
M_\sigma&=&M\big(\cos\lambda+\sigma\sin\lambda\big),\\
\mu_\sigma&=&\frac{2M^2+\tilde{M}^2\cos^2\lambda\big[3+2\sigma\sin(2\lambda)-\cos(2\lambda)\big]}{E_{F_\perp}},\\
\upsilon_{F_{\perp,\sigma}}&=&\frac{2M\tilde{M}\cos\lambda\big[1+\sigma\sin(2\lambda)\big]}{E_{F_\perp}^2}\upsilon_{F_\perp}\equiv
\frac{2\theta M^2\cos\lambda\big[1+\sigma\sin(2\lambda)\big]}{E_{F_\perp}^2}\upsilon_{F_\perp}\,.
\eea

\noi From the structure of the above relations, we infer that the perturbative approach considered here is valid only as long as the following energy hierarchy holds $|M|\ll E_{F_\perp}\ll E_F$.

Remarkably, each block Hamiltonian coincides with the Hamiltonian describing a spin-$\frac{1}{2}$ hole-like Rashba 2DEG in the presence of proximity induced superconductivity, spatial anisotropy, and an out-of-plane Zeeman field~\cite{SauGeneric}. Such a correspondence becomes established by mapping the movers space in the present analysis to the spin degrees of freedom of the holes in the 2DEG. These Hamiltonians belong to symmetry class D and allow for chiral Majorana modes at termination edges or other types of linear defects, e.g., domain walls. Given the structure of the above Hamiltonians we expect to obtain up to a single chiral Majorana edge mode per $\sigma$ block. Since class D is characterized by a $\mathbb{Z}$ invariant, the Majorana excitations arising from each block remain robust against sufficiently weak inter-block mixing. Note also that for the special case of $\lambda=\pi/4$, only the $\sigma=+$ block experiences magnetic order and thus we can have only up to a single chiral Majorana mode per edge.

The emergence of chiral Majorana edge modes can be associated with the closing and reopening of the bulk energy gap. For a nonzero $\tilde{M}$, gap closings can only happen at the high-symmetry point $\bm{p}=\bm{0}$, where the effective SOC term $\upsilon_Fp_1\rho_2-\upsilon_{F_{\perp,\sigma}}p_2\rho_1$ vanishes. The gap closes at $\bm{p}=\bm{0}$ when the conditions $|M_\sigma|=\sqrt{\Delta^2+\mu_\sigma^2}$ are satisfied. At this point, it is interesting to compare these findings with the behavior of the system when $\tilde{M}=0$ and the magnetic order is of the SWC$_4$ type, which is not skyrmionic. As first shown in Ref.~\citenum{SteffensenPRR}, in the latter scenario the bulk energy spectrum is nodal, with pairs of nodes appearing at points of the form $(p_1=0,\pm p_{\rm node})$, as shown see~\ref{fig:Figure1}(d). This becomes transparent by noticing that now the previous gap closing condition $|M_\sigma|=\sqrt{\Delta^2+\mu_\sigma^2}$ becomes replaced by $|M_\sigma|=\sqrt{\Delta^2+(p_2^2/2m_\perp-\mu_\sigma)^2}$. Hence, the band touching obtained for the class D Hamiltonian at $\bm{p}=\bm{0}$ propagates to nonzero $p_2$ when $\tilde{M}$ is absent and chiral symmetry emerges. For a system with chiral symmetry, the system does not support chiral Majorana edge modes but instead Majorana flat bands~\cite{SteffensenPRR}. Therefore, we find that the generation of a nonzero $\tilde{M}$ by means of MSC is instrumental for gapping out the system and endowing the Majorana flat bands with a chiral dispersion.

\section{Self-Consistent Approach to Magnetic Skyrmion Catalysis}\label{Sec:SectionVI}

Having obtained a low-energy effective model describing the superconductor in the topologically nontrivial regime, we here investigate the spontaneous emergence of magnetism. We work in the weak-coupling limit, in which, the pairing and magnetic gaps are sufficiently smaller than the Fermi energy $E_{F_\perp}$ dictating transverse motion. In this limit, the low-energy model can be linearized with respect to the momenta and the chemical potential can be discarded. Therefore, we end up with the approximate Dirac-type energy dispersions of the form $\pm E_{\sigma,\pm}(\bm{p})$, where we introduced:
\begin{align}
E_{\sigma,\pm}(\bm{p})=\sqrt{\big(\upsilon_Fp_1\big)^2+\zeta_\sigma^2\big(\upsilon_{F_\perp}p_2\big)^2+M_{\sigma,\pm}^2}\,,
\end{align}

\noi that we express in a compact fashion by correspondingly introducing the effective gaps and velocity prefactors:
\begin{align}
M_{\sigma,\pm}=M_\sigma\pm\Delta\qquad{\rm and}\qquad\zeta_\sigma=2\theta \cos\lambda\big[1+\sigma\sin(2\lambda)\big]\left(\frac{M}{E_{F_\perp}}\right)^2\,.
\end{align}

To determine the profile of the Sk-SWC$_4$ phase self-consistently we essentially need to minimize the free energy of the system with respect to $M$ and $\lambda$. By phenomenologically invoking an attractive interaction in the magnetic channel of strength $U$, we obtain the energy per area of the system at zero temperature:
\begin{align}
E=\frac{|\bm{M}_1|^2+|\bm{M}_2|^2}{U}-\sum_{\sigma,\nu=\pm}\int_{-E_{F_\perp}}^{+E_{F_\perp}}\frac{d\epsilon_2}{2\pi\upsilon_{F_\perp}}\int_{-E_F}^{+E_F}\frac{d\epsilon_1}{2\pi\upsilon_F}\sqrt{\epsilon_1^2+\zeta_\sigma ^2\epsilon_2^2+M_{\sigma,\nu}^2}\,,
\end{align}

\noi which only takes contributions from the negative/occupied states. Note that the above form was obtained after switching from momentum to energy variables. Further, we introduced the energy cutoffs defined by the Fermi energies $E_F$ and $E_{F_\perp}$. We remark that the factor of $1/2$ which is required to be added in order to avoid double-counting in the BdG formalism does not appear in the above, because it is compensated by a factor of $2$ which arises from accounting for the second pair of nested Fermi lines appearing along the $p_2$ axis.

To proceed, we replace $\bm{M}_{1,2}$ with the expressions in Eq.~\eqref{eq:M12} and carry out the integral over $\epsilon_1$. To make the expression more transparent, we introduce the density of states of the system in the absence of superconductivity $\nu_0=m_2/2\pi^2$ and we end up with the following expression for the rescaled energy per area $\bar{E}=E/\nu_0$:
\begin{align}
\bar{E}=\frac{2M^2}{\bar{U}}-\sum_{\sigma,\nu=\pm}\int_{-E_{F_\perp}}^{+E_{F_\perp}}\frac{d\epsilon_2}{4E_{F_\perp}}\ph
\left\{E_F\sqrt{E_F^2+\zeta_\sigma^2\epsilon_2^2+M_{\sigma,\nu}^2}+\Big(\zeta_\sigma^2\epsilon_2^2+M_{\sigma,\nu}^2\Big)\ln\left(\frac{\sqrt{E_F^2+\zeta_\sigma^2\epsilon_2^2+M_{\sigma,\nu}^2}+E_F}{\sqrt{\zeta_\sigma^2\epsilon_2^2+M_{\sigma,\nu}^2}}\right)\right\},
\end{align}

\noi where we replaced $\upsilon_F$ in terms of $E_{F_\perp}$ and introduced the rescaled interaction strength $\bar{U}=\nu_0U$.

\subsection{Solution for $\theta=\Delta=0$}

In order to tackle the problem, it is helpful to first better understand the behavior of $M$ when $\theta=\Delta=0$, i.e, in the absence of ground state flux and proximity induced superconductivity. Essentially, this correponds to the situation of a spin density wave arising from nested flat Fermi lines. In this case, we obtain the simplified expression:
\begin{align}
\bar{E}_{\theta=\Delta=0}=\frac{2M^2}{\bar{U}}-\sum_{\sigma=\pm}\ph
\left\{E_F\sqrt{E_F^2+M_\sigma^2}+M_\sigma^2\ln\left(\frac{\sqrt{E_F^2+M_\sigma^2}+E_F}{|M_\sigma|}\right)\right\}.
\end{align}

\noi Extremizing $\bar{E}_{\theta=\Delta=0}$ with respect to $M$ throught $\partial \bar{E}_{\theta=\Delta=0}/\partial M=0$, leads for $M\neq0$ to the condition:
\begin{align}
\frac{4M}{\bar{U}}=\sum_{\sigma=\pm}\frac{\partial M_\sigma}{\partial M}2M_\sigma\ln\left(\frac{\sqrt{E_F^2+M_\sigma^2}+E_F}{|M_\sigma|}\right)\Rightarrow
\frac{2}{\bar{U}}=\sum_{\sigma=\pm}\big(\cos\lambda+\sigma\sin\lambda\big)^2\ln\left(\frac{\sqrt{E_F^2+M_\sigma^2}+E_F}{|M_\sigma|}\right)\,.
\label{eq:Selfie1D}
\end{align}

\noi We plug the above condition in $\bar{E}_{\theta=\Delta=0}$ and find that:
\begin{align}
\bar{E}_{\theta=\Delta=0}^{\rm min}=-E_F\sum_{\sigma=\pm}\sqrt{E_F^2+M_\sigma^2}=-E_F\sum_{\sigma=\pm}\sqrt{E_F^2+M^2+\sigma M^2\sin(2\lambda)}\,,
\end{align}

\noi where the value of $M$ entering in the above equations is obtained from the solution of the self-consistency Eq.~\eqref{eq:Selfie1D}.

Note that also the value of $\lambda$ is to be found by means of extremizing the energy functional. However, one needs to be cautious here. The present analysis neglects energy contributions from momentum space regions which are away from the flat bands. Omitting these can affect the minimization outcome for $\lambda$, since this degree of freedom is sensitive to couplings between $\bm{M}_1$ and $\bm{M}_2$. Nonetheless, extremizing the energy functional at hand with respect to $\lambda$ for $M\neq0$, yields that the following condition needs to be satisfied:
\begin{align}
\cos(2\lambda)\sum_{\sigma=\pm}\sigma\frac{1}{\sqrt{E_F^2+M^2+\sigma M^2\sin(2\lambda)}}=0\Rightarrow \lambda=\left\{0,\pm\frac{\pi}{4},\pm\frac{\pi}{2},\pi\right\}\,.
\end{align}

\noi While one finds that $\lambda=\{0,\pm\pi/2,\pi\}$ is the solution that leads to the minimum and thus does not allow for the stabilization of the SWC$_4$ phase within this low-energy model, in the limit $M\ll E_F$ which is of interest here, the phases with $\lambda=\pm\pi/4$ and $\lambda=\{0,\pm\pi/2,\pi\}$ are practically degenerate. Therefore, inferring the value for $\lambda$ by only accounting for the nested Fermi segments' contribution to the energy seems inconclusive on whether the SWC$_4$ phase becomes stabilized, since the abovementioned neglected contributions to the energy can modify this picture. In fact, by assuming that $U\ll E_F$ and thus $M\ll E_F$, one finds from Eq.~\eqref{eq:Selfie1D} that for $\lambda=\{0,\pm\pi/2,\pi\}$ we have $|M_\sigma|=M$ and the latter satisfies $M\approx 2E_Fe^{-1/\bar{U}}$. In contrast, for $\lambda=\pm\pi/4$ only one of the $M_\sigma$ is nonzero and equal to $\sqrt{2}M$. In this case, the self-consistency equations yields $M\approx \sqrt{2}E_Fe^{-1/\bar{U}}$.

Given the fact that within the low-energy picture adopted here it is not possible to settle whether the SWC$_4$ phase is thermodynamically favored to emerge, in the remainder, we will simply assume that it is indeed stabilized for some value of $\lambda$ that we nevertheless cannot have information about. Hence, in the following, we take for granted that the SWC$_4$ phase is stabilized and we solely examine the behavior of $M$ using the self-consistent approach for an ad hoc selected value of $\lambda$. Assuming that SWC$_4$ is a plausible physical scenario is also supported by theoretical predictions regarding its emergence in doped Fe-based superconductors~\cite{Christensen_18}. In addition, we remind the reader that our findings mainly have a qualitative character and the precise value of $\lambda$ is not crucial. In fact, the analysis of the upcoming paragraph shows that a similar trend emerges for different values of $\lambda$, thus choosing a specific value for $\lambda$ does not harm the generality of the conclusions arising from our investigation.

\subsection{Solution for $\theta\neq0$ and $\Delta\neq0$}

Equipped with the above valuable insights, we now proceed with attacking the self-consistency problem for generally nonzero $\theta$ and $\Delta$. Since in the present case it is difficult to obtain analytical results, we minimize the energy functional numerically. Given the form of the energy functional it is more convenient to first carry out all the integrals. This procedure yields the following closed form expression for the rescaled energy functional:
\begin{align}
\bar{E}=\frac{2M^2}{\bar{U}}-\frac{1}{3}\sum_{\sigma,\nu=\pm}\ph
\left\{E_F\sqrt{E_F^2+\zeta_\sigma^2E_{F_\perp}^2+M_{\sigma,\nu}^2}+\frac{\zeta_\sigma^2E_{F_\perp}^2+3M_{\sigma,\nu}^2}{2}\ln\left(\frac{\sqrt{E_F^2+\zeta_\sigma^2E_{F_\perp}^2+M_{\sigma,\nu}^2}+E_F}{\sqrt{\zeta_\sigma^2E_{F_\perp}^2+M_{\sigma,\nu}^2}}\right)\right.\qquad\qquad\no\\
-\left.\frac{M_{\sigma,\nu}^3}{\zeta_\sigma E_{F_\perp}}
\tan^{-1}\left(\frac{\zeta_\sigma E_{F_\perp}E_F}{M_{\sigma,\nu}\sqrt{E_F^2+\zeta_\sigma^2E_{F_\perp}^2+M_{\sigma,\nu}^2}}\right)+\frac{E_F\big(E_F^2-3M_{\sigma,\nu}^2\big)}{2\zeta_\sigma E_{F_\perp}}\ln\left(\frac{\sqrt{E_F^2+\zeta_\sigma^2E_{F_\perp}^2+M_{\sigma,\nu}^2}+\zeta_\sigma E_{F_\perp}}{\sqrt{E_F^2+M_{\sigma,\nu}^2}}\right)
\right\}.
\end{align}

\begin{figure}[t!]
\begin{center}
\begin{tabular}{c}
\includegraphics[width=0.89\textwidth]{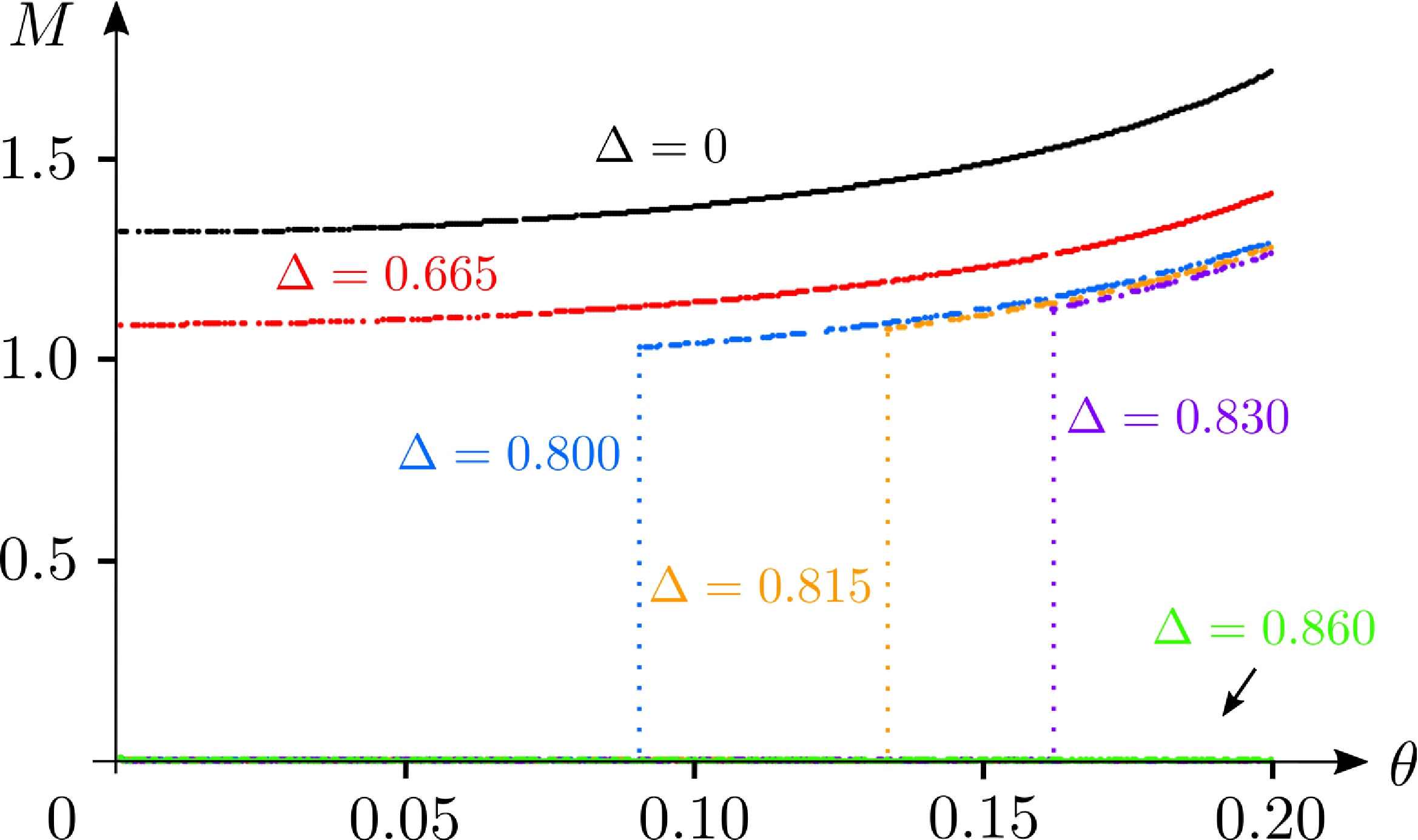}\vspace{0.1in}
\end{tabular}
\end{center}
\caption{\label{fig:Figure2}
Self-consistent magnetization modulus $M$ as a function of the ground state flux $\theta$. We find that $M$ becomes enhanced upon increasing $\theta$, thus revealing the ``catalytic" role that flux plays in stabilizing a Sk-SWC$_4$ phase when starting from the nonskyrmionic SWC$_4$ phase at zero flux. Even more, tuning the flux allows compensating the suppression of $M$ which is brought about from the pairing gap. Since magnetism and spin-singlet superconductivity are antagonistic, superconductivity generally tends to suppress magnetism. This general rule is also confirmed in our findings. Indeed, for sufficiently small values of the flux, increasing $\Delta$ decreases or fully eliminates magnetism. However, the SK-SWC$_4$ phase can re-appear upon increasing the flux. Note that the perpendicular lines are guides to the eye, showing the discontinuous re-entrance of magnetism starting from a pure nonmagnetic superconducting ground state. These results demonstrate that the MSC can enhance the robustness of the topological superconductor. This is because magnetism can be controlled by the flux and the magnetization modulus can be rendered larger than the pairing gap, which is a requirement to access the topological phase. For the numerics, we used the parameter values $E_F=10$, $E_{F_\perp}=1$, $\bar{U}=0.4$, and $\lambda=\pi/3$.}
\end{figure}

Instead of determining $M$ using the self-consistency condition, we here directly minimize the rescaled energy functional shown above in a finite domain for $M$ which fully includes the regions of negative energy. This minimization process leads to the results shown in Fig.~\ref{fig:Figure2}. There, we indeed confirm that the violation of flux and the concomitant emergence of a nonzero $\theta$ favor the stabilization of the magnetic order since $M$ increases. In fact, such an enhancement trend is a typical characteristic of a catalysis mechanism, as it has been also shown in previous works investigating the magnetic-fied induction of $d+id$ chiral superconductors~\cite{Laughlin} and particle-hole condensates~\cite{BalatskyCDDW,KotetesCondensates,KotetesNernst,KotetesPhilo}. Moreover, for a fixed value for $\theta$, we confirm the suppression of the magnetic order upon increasing the strength of the spin-singlet pairing gap $\Delta$. This is in accordance with the findings of Ref.~\citenum{Mendler} on the competition between conventional superconductivity and skyrmion MTCs. Remarkably, here the magnetic order exhibits re-entrant behavior upon compensating $\Delta$ by increasing $\theta$. Therefore, we conclude that the skyrmion MTC shows wide windows of robustness even in the presence of a spin-singlet pairing gap. Even more, $M$ can be larger than $\Delta$ which is a requirement for entering the topologically nontrivial superconducting regime. Finally, we find clear evidence for the mechanism of MSC, since increasing $\theta$ leads to an increase of $M$ and the further stabilization of the Sk-SWC$_4$ phase. Hence, the Sk-SWC$_4$ phase is more robust than the SWC$_4$ ground state.

\section{Conclusions and Outlook}\label{Sec:SectionVII}

In this work we demonstrated how the mechanism of magnetic skyrmion catalysis (MSC) can lead to topological superconductivity with protected chiral Majorana edge modes. We particularly focused on a concrete example, i.e., a skyrmionic spin whirl crystal (Sk-SWC$_4$) phase, and derived an effective low-energy theory which reveals the mapping to the celebrated model for topological superconductivity in noncentrosymmetric superconductors in the additional presence of an out-of-plane spin splitting~\cite{FujimotoSato,SauGeneric}. Our analysis provides support to the claimed catalytic nature of the mechanism proposed here, since we show that in the presence of a ground state flux, the Sk-SWC$_4$ becomes stabilized and its robustness is enhanced. This result is obtained using a self-consistent approach for the magnetization modulus using the low-energy model derived earlier.

Concerning the experimental realization of MSC, we remark that the SWC$_4$ has been theo\-re\-ti\-cal\-ly predicted~\cite{Christensen_18} for hole-doped BaFe$_2$As$_2$, while the noncollinear spin-vortex phase, which can also be rendered skyrmionic in the presence of flux~\cite{Huang}, has been already found in CaKFe$_4$As$_4$~\cite{SVCexp}. Hence, Fe-based superconductors with coexisting magnetism and super\-con\-duc\-ti\-vi\-ty~\cite{ni08a,nandi10a,avci11,johrendt11,klauss15,wang16a} appear promising to exhibit intrinsic chiral superconductivity once flux emerges. Another class of systems that can potentially exhibit intrinsic topological superconductivity is the recently discovered family of Kagome superconductors~\cite{UCDW_KVSb,muonGraf,muonKVSb,OpticalDetecCDW}. These harbor superconductivity and a flux phase~\cite{AHE_CDW_CVSb,HiddenFluxPhaseCVSb,UCDW_RbVSb} which leads to a nonzero Berry curvature in the energy bands~\cite{ChiralFluxKagome,Nandkishore}. Thus, the possible additional emergence of magnetism in these systems opens perspectives for realizing chiral Majorana edge modes.

\acknowledgements PK acknowledges funding from the National Na\-tu\-ral Science Foundation of China (Grant No.~12250610194).

\bibliography{RefsSPIEman}

\bibliographystyle{spiebib}

\end{document}